\documentclass[twocolumn]{aastex62}
\usepackage{natbib}
\graphicspath{{./}{../pictures/}}

\received{}
\revised{}
\accepted{}
\submitjournal{ApJL}

\shorttitle{Double flares in NSCs}
\shortauthors{Ryu et al.}
\usepackage{graphicx}	% Including figure files
\usepackage{amsmath}	% Advanced maths commands
\usepackage{amssymb}	% Extra maths symbols
\usepackage{cases}
\usepackage{array,multirow}
\usepackage{upgreek}
\usepackage{textcomp}

\usepackage{hyperref}
\usepackage[referable]{threeparttablex}
\usepackage{booktabs, dcolumn}
\makeatletter
\newcommand*{\rom}[1]{\expandafter\@slowromancap\romannumeral #1@}
\makeatother

\newcommand{\beq}{\begin{equation}}
\newcommand{\eeq}{\end{equation}}
\newcommand{\simlt}{\mathrel{\hbox{\rlap{\hbox{\lower4pt\hbox{$\sim$}}}\hbox{$<$}}}}
\newcommand{\simgt}{\mathrel{\hbox{\rlap{\hbox{\lower4pt\hbox{$\sim$}}}\hbox{$>$}}}}

\newcommand{\Msol}{\,\mathrm{M}_\odot}
\newcommand{\Rsol}{\,\mathrm{R}_\odot}

\newcommand{\yr}{\,\mathrm{yr}}
\newcommand{\pergal}{\,\mathrm{gal}^{-1}}
\newcommand{\pc}{\,\mathrm{pc}}

\newcommand{\Mbh}{M_{\bullet}}
\newcommand{\mbh}{m_{\bullet}}
\newcommand{\ms}{m_{\mathrm{\star}}}
\newcommand{\rs}{r_{\mathrm{\star}}}

\def\apj{ApJ}
\def\mnras{M.N.R.A.S.}
\def\aap{A\&A}
\def\nat{Nat.}

\def\araa{Ann. Rev. A\&A}

\def\aapr{A\&A Rev.}

\def\prd{prd}

\usepackage[normalem]{ulem}

\begin{document}

\title{TIDAL DISRUPTION ENCORES}

\correspondingauthor{Taeho Ryu}
\email{tryu@mpa-garching.mpg.de}

\author[0000-0002-0786-7307]{Taeho Ryu}
\affil{The Max Planck Institute for Astrophysics, Karl-Schwarzschild-Str. 1, Garching, 85748, Germany}
\affiliation{Physics and Astronomy Department, Johns Hopkins University, Baltimore, MD 21218, USA}

\author[0000-0002-3635-5677]{Rosalba Perna}
\affil{Department of Physics and Astronomy, Stony Brook University, Stony Brook, NY 11794-3800, USA}
\affiliation{Center for Computational Astrophysics, Flatiron Institute, New York, NY 10010, USA}

\author{Matteo Cantiello}
\affiliation{Center for Computational Astrophysics, Flatiron Institute, New York, NY 10010, USA}

\begin{abstract}
Nuclear star clusters (NSCs), made up of a dense concentrations of stars and the compact objects they leave behind, are ubiquitous in the central regions of galaxies, surrounding the central supermassive black hole (SMBH). Close interactions between stars and stellar-mass black holes (sBH) lead to tidal disruption events (TDEs). We uncover an interesting new phenomenon: For a subset of these, the unbound debris (to the sBH) remain bound to the SMBH, accreting at a later time, and thus giving rise to a  second flare. 
We compute the rate of such events, and find them ranging within $10^{-6}$ - $10^{-3}$~yr$^{-1}$gal$^{-1}$ for SMBH mass $
\simeq 10^{6}-10^{9}\Msol$. Time delays between the two flares spread over a wide range,
from less than a year to hundreds of years.
The temporal evolution of the light curves of the second flare can vary
between the standard $t^{-5/3}$ power-law to much steeper decays,  providing a natural explanation for observed light curves in tension with the classical TDE model. Our predictions have implications for learning about NSC properties and calibrating its sBH population. Some double flares may be electromagnetic counterparts to LISA Extreme-Mass-Ratio-Inspiral (EMRI)
sources. Another important implication  is the possible existence of  TDE-like events in very massive SMBHs, where TDEs are not expected. Such flares can affect spin measurements relying on TDEs in the upper SMBH range.
\end{abstract}


\section{Introduction} \label{sec:intro}

Most galactic nuclei are characterized by an
extremely high density of stars. These nuclear star clusters (NSCs),
which are unrivaled in luminosity compared to any other type of 
star clusters, possess the densest known stellar densities (see \cite{Neumayer2020} for an extensive review).
Their steep mass profiles, ranging from $\rho(r)\propto r^{-1}$ to
$r^{-3}$, result in densities which are $\sim 10^6 \Msol\pc^{-3}$
at radial distances $R\sim 0.1\pc$.
Their occupation fraction appears to vary with
both galaxy mass and galaxy type, reaching $\gtrsim 80\%$ in $\sim 10^9\Msol$ early type galaxies \citep{Sanchez2019}, but it declines at both the lower and the higher-mass end. The study of NSCs is an important one in galaxy formation, their formation being tied to both the growth of galaxies as well as to their central supermassive black hole (SMBH). 

The very high densities of NSCs, which contain a fraction $\sim 1\%$ of their mass in stellar-mass black holes (sBH, \citealt{BahcallWolf1976}), are very conducive to close encounters among their constituents. In particular, close encounters between sBHs and stars in the NSC can lead to tidal disruption events (TDEs), detectable by their high luminosities. The occurrence of these events in NSCs is especially important, in that it has been shown to help calibrate the number of sBH binaries in these systems \citep{Fragione2021}.

\begin{figure*}
    \centering
    \includegraphics[width=5.9cm,trim={0.7cm 0 0 0},clip]{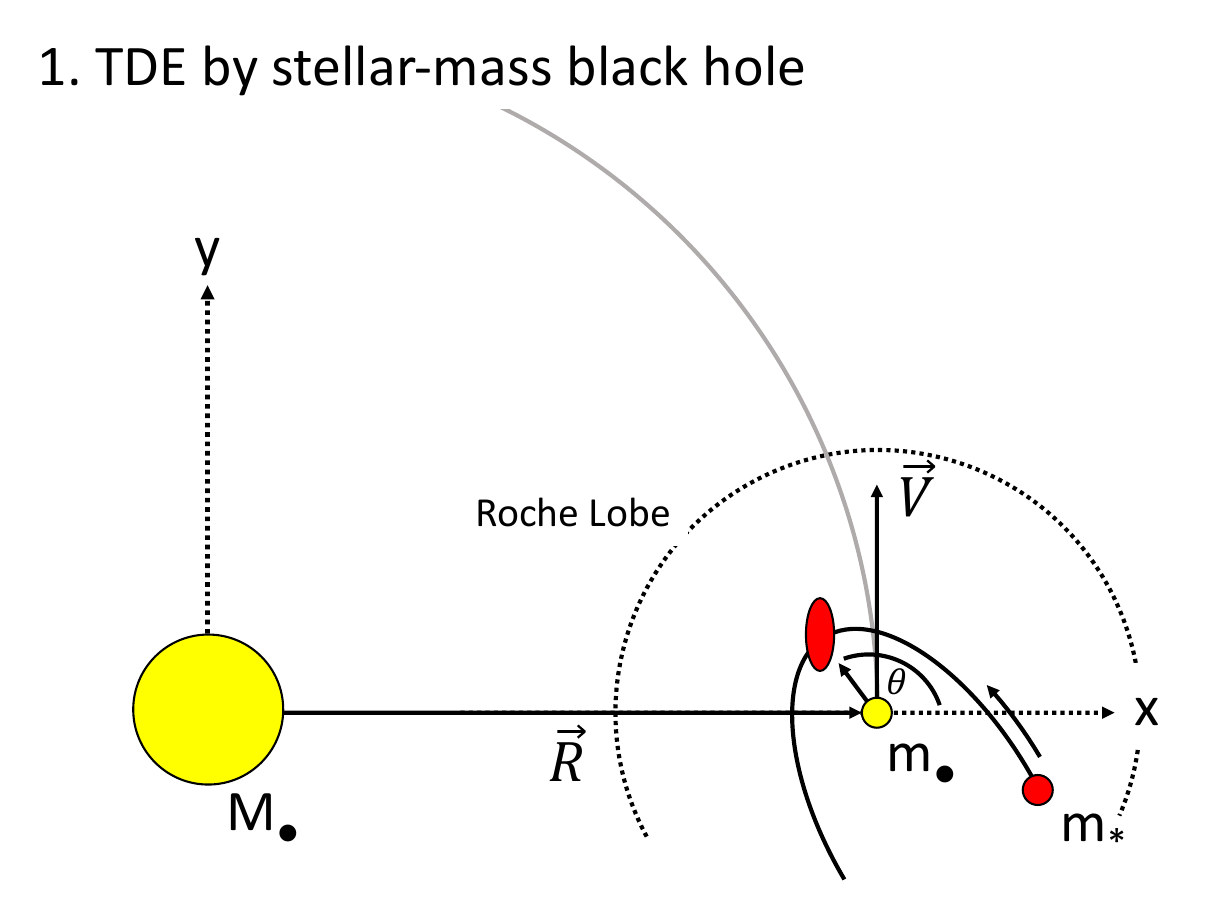}
    \includegraphics[width=5.9cm,trim={0.7cm 0 0 0},clip]{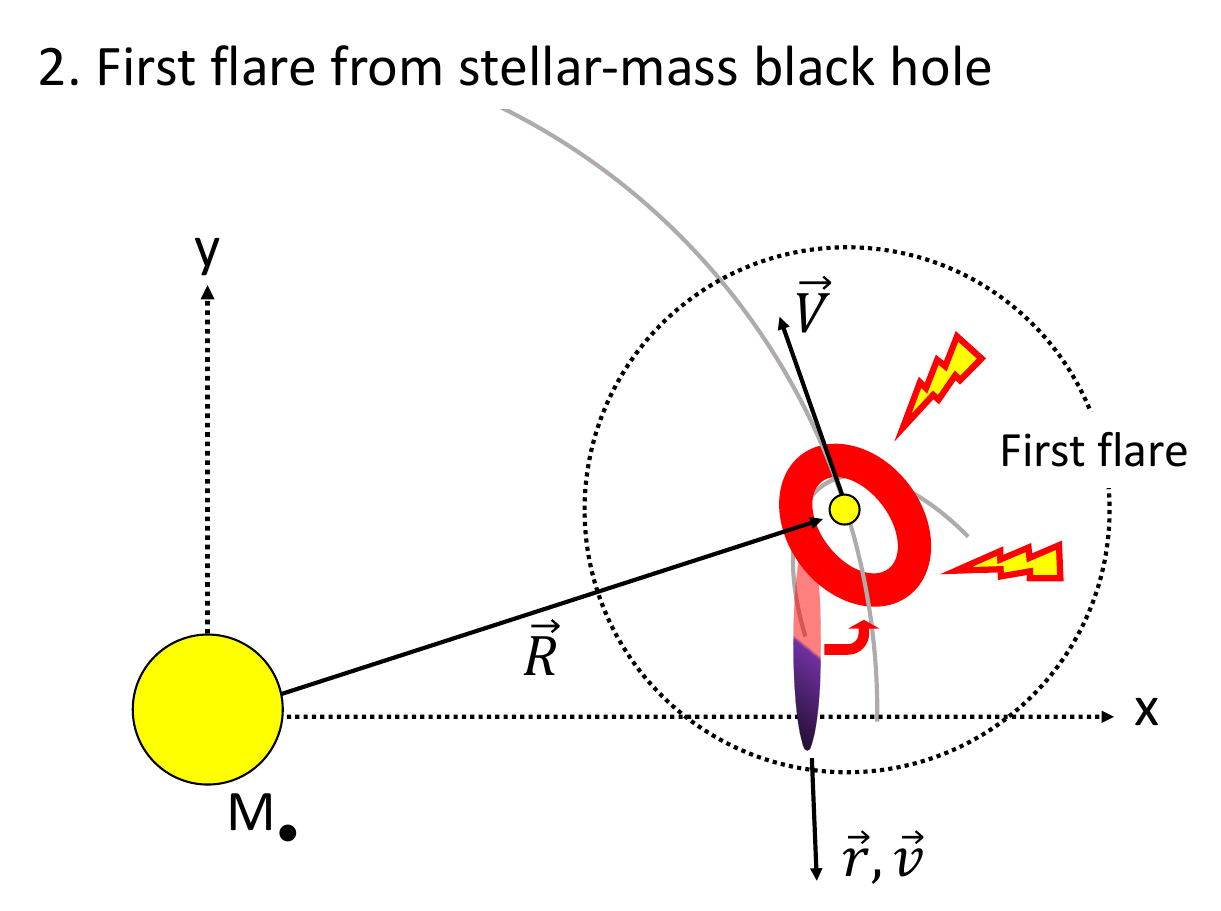}
    \includegraphics[width=5.9cm,trim={0.7cm 0 0 0},clip]{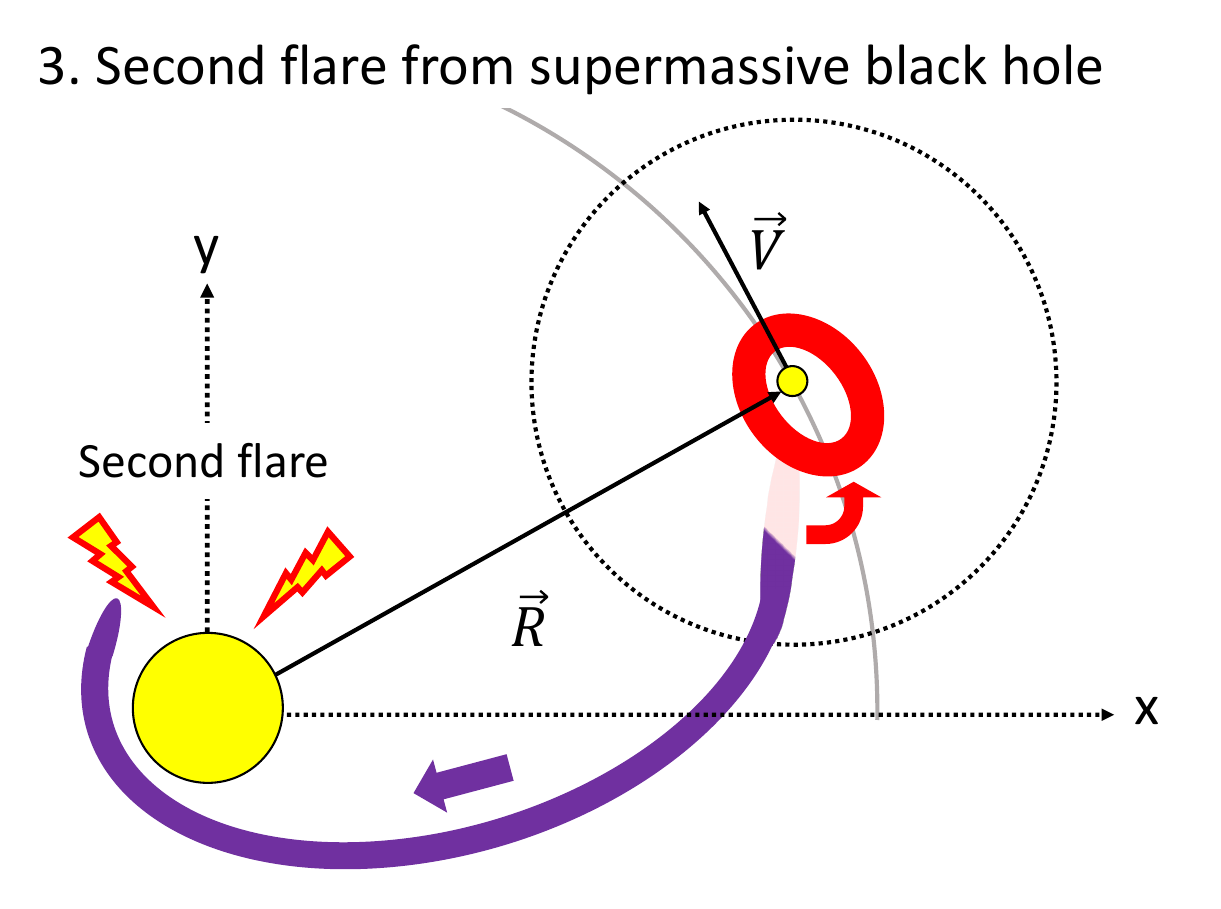} 
    \caption{\label{fig:schematic1} Schematic diagram for a double-flare event in a 2D configuration. While an sBH (small yellow circle) is orbiting around a SMBH (large yellow circle), it tidally destroys a star (red circle) (\textit{left}). The angle of the incoming star's orbit at disruption relative to the line connecting the sBH and SMBH is $\theta$.  The debris bound to the sBH returns and creates the first flare (\textit{middle}). The debris, originally unbound from the sBH, spills out of its Roche Lobe. If the kinetic energy of the spilled debris is not sufficiently high, it  remains bound to the SMBH and falls into it, creating the second flare (\textit{right}). \textit{The diagram is not to scale.}   }
\end{figure*}

Here we note an interesting new phenomenon. When a star in the NSC undergoes a TDE by an sBH \citep[][]{Perets2016}, some fraction of the debris remain bound and accrete to the sBH, while the rest gets unbound. However, there is a range of locations of the sBH for which these unbound debris, while escaping from the sBH,  remain bound to the SMBH, to which they hence return, causing a second, longer flare. In this paper we quantify the conditions under which the sBH-unbound debris remain SMBH-bound, as a function of the relevant variables of the problem, that is the sBH location, the SMBH mass, and the debris ejection angle. We compute the fallback rates for a number of representative cases, discovering a variety of resulting temporal slopes which can yield 
a varied range of electromagnetic transients.
We finally compute the event rates per galaxy of these double flaring events, showing that they can be as high as the standard TDE rates, depending on the SMBH mass.

Our paper is organized as follows: \S\ref{sec:orbit}
illustrates the geometry of the problem and the orbital properties of the debris. Fallback rates are presented in \S\ref{sec:fallbackrate}, together with a discussion of the electromagnetic signatures of the double flaring event. We compute the rate in \S\ref{sec:rate}. We summarize and discuss the implications of our results in \S\ref{sec:conclusion}.

\section{Orbit of debris around a SMBH - sBH system}\label{sec:orbit}

We consider a TDE where a $\ms=1\Msol$ and $\rs=1\Rsol$ main-sequence star is tidally disrupted at the tidal radius $r_{\rm t}=(\mbh/\ms)^{1/3}\rs$ by a sBH with mass $\mbh$ moving on a circular orbit around a SMBH at a distance $R$. For simplicity, we assume that the original orbit of the star relative to the sBH is parabolic and the sBH-SMBH orbit aligns with the orbit of the disrupted star around the sBH (thus in the same orbital plane). 

\subsection{The orbit of debris originally bound to the sBH}

Most of the debris originally bound to the sBH would return to the sBH, creating the first flare. However, some fraction of the originally bound debris is still energetic enough to go beyond the Roche Lobe of the sBH, subsequently falling into the SMBH. As an order of magnitude estimate, the minimum energy required to reach the Roche Lobe (RL) radius of the sBH $r_{\rm RL}\simeq (\mbh/3\Mbh)^{1/3}R$ is $\simeq -G\mbh/[2 r_{\rm RL}]$. Assuming a top-hat energy distribution ($dm/d\epsilon\simeq \ms/ 2\Delta\epsilon$ where $\epsilon$ is the debris' orbital energy relative to the sBH and $\Delta\epsilon=G\mbh \rs/r_{\rm t}^{2}$), the fraction of originally bound debris mass which can escape the RL of the sBH, relative to the stellar mass, can be roughly estimated as
\begin{align}\label{eq:ratio}
 f_{>r_{\rm RL}}&\simeq \frac{G\mbh/[2 r_{\rm RL}]}{\Delta\epsilon},\nonumber\\
                &\simeq 10^{-4} \left(\frac{\Mbh}{10^{6}\Msol}\right)^{1/3}\left(\frac{\mbh}{10\Msol}\right)^{1/3}\left(\frac{\ms}{1\Msol}\right)^{-2/3}\nonumber\\
                &\times\left(\frac{\rs}{1\Rsol}\right)
                   \left(\frac{R}{10^{-2}\pc}\right)^{-1}.
\end{align}
This estimate suggests that the bound debris is mostly contained within the RL of the sBH, and only a small fraction may flow towards the SMBH for very massive SMBHs. Therefore, in the following we will assume that the entire originally bound debris stays within the RL of the sBH and is responsible for creating the first flare. 

\subsection{The orbit of the originally unbound debris}

\begin{figure*}
    \centering
    \includegraphics[width=8.5cm]{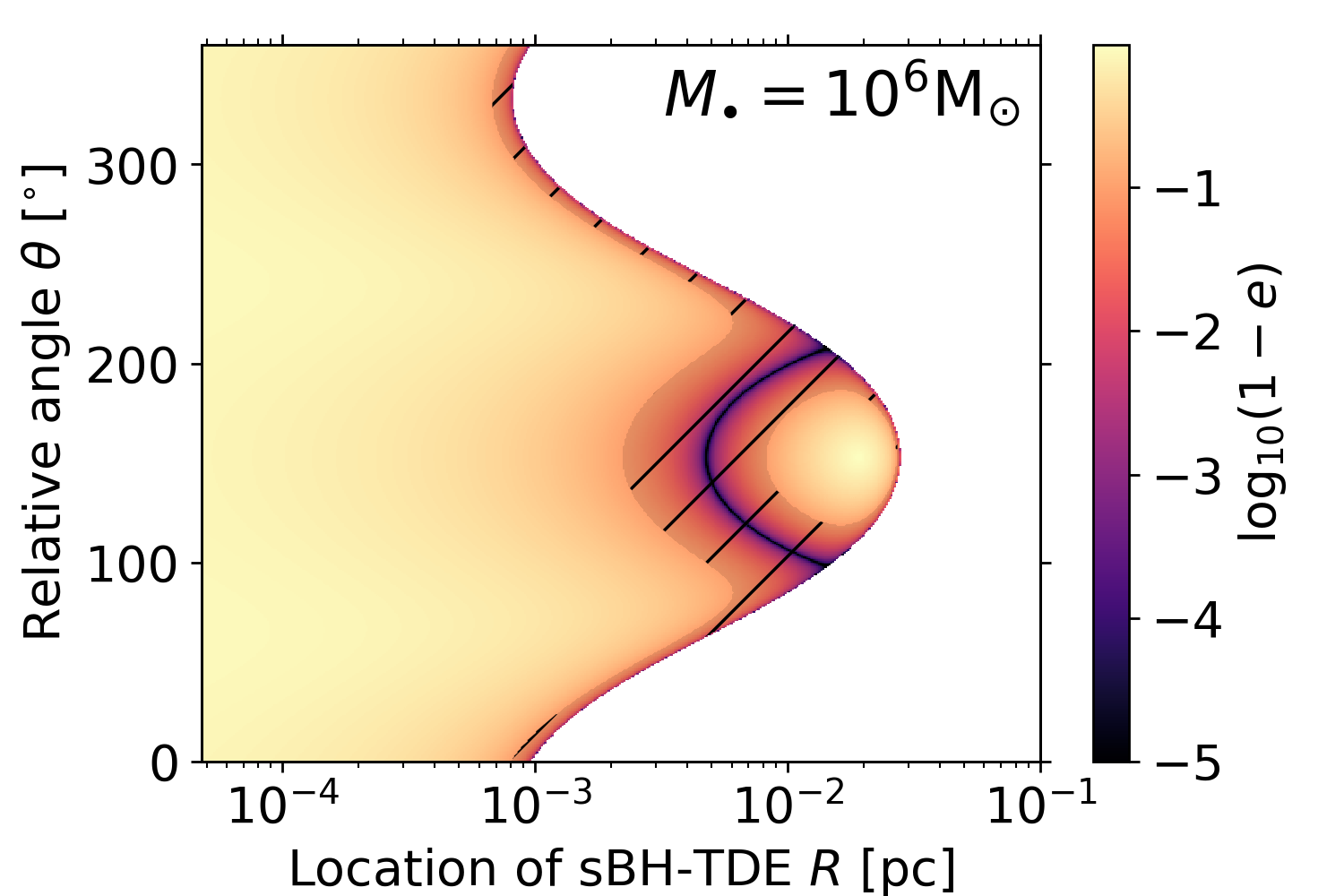}  
    \includegraphics[width=8.5cm]{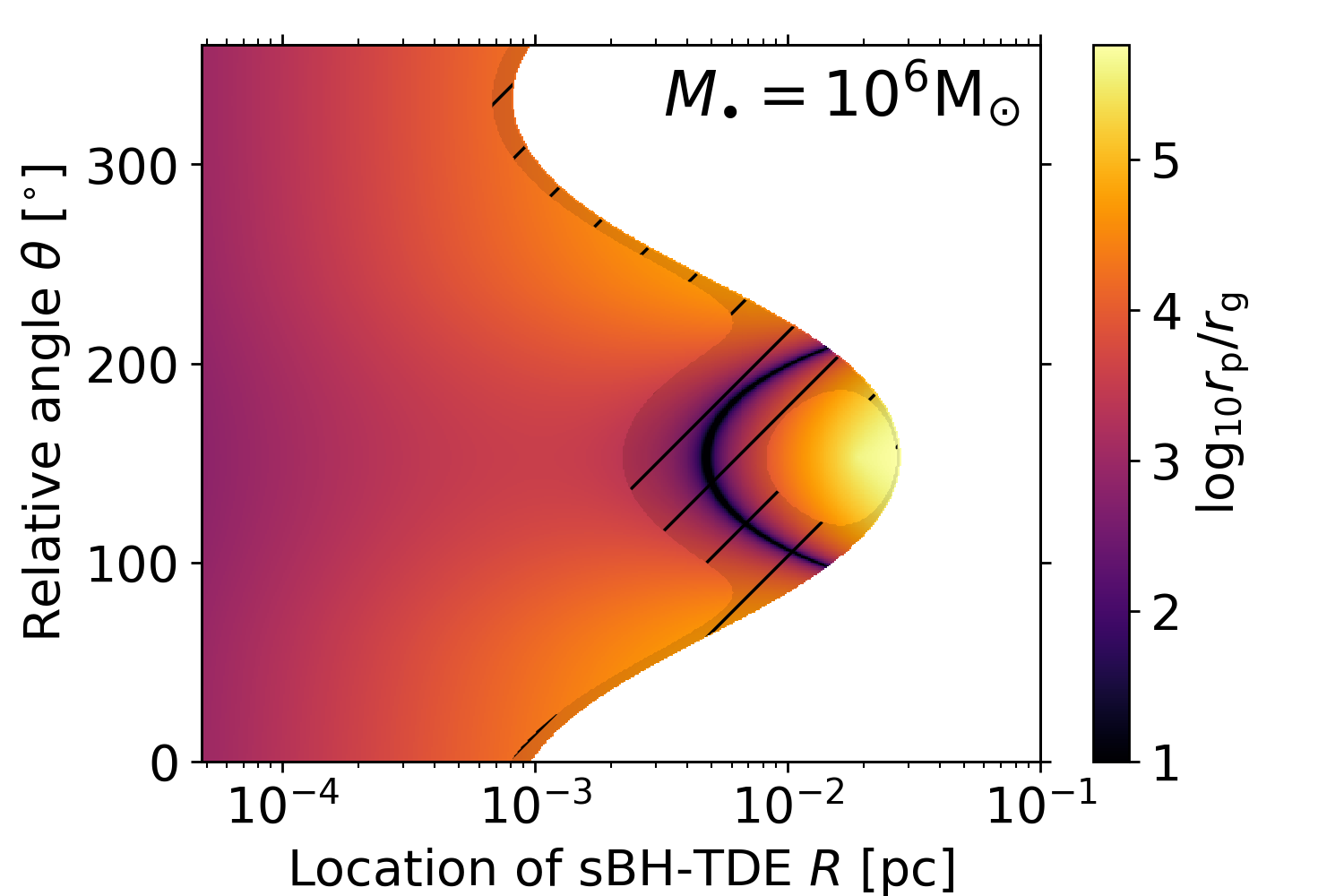}      
    \caption{\label{fig:orbit_hyperbolic} Eccentricity $e$ (\textit{left}) and pericenter distance $r_{\rm p}$ (\textit{right}) of the ``spilled'' debris unbound from the sBH but bound to the $10^{6}\Msol$ SMBH. Those quantities are measured for the debris' orbit relative to the SMBH.  In both panels, the hatched regions indicate highly eccentric debris ($e>0.9$). The region with a white background indicates debris unbound from the SMBH. The binding energy of the spilled debris relative to the SMBH is determined by its kinetic energy at the point it escapes the Roche-Lobe around the sBH at a given location. Debris moving in the opposite direction to the sBH ($\theta\simeq 180^{\circ}$) tends to have less kinetic energy and smaller angular momentum, resulting in a more bound and more eccentric orbit around the SMBH. As the distance of the sBH-TDEs from the SMBH gets smaller, the kinetic energy of the sBH is more dominant with respect to the debris' kinetic energy, leading to the spilled debris' orbit becoming more similar to that of the sBH.  }
\end{figure*}

We first consider the orbit of the most energetic ejecta relative to the sBH. Within the RL of the sBH, its position and velocity in the frame of the sBH can be described by an hyperbolic orbit with $\epsilon=\Delta \epsilon$ and angular momentum such that the pericenter distance is the tidal radius $r_{\rm t}$ from the sBH. Here, we introduce the angle $\theta$ between the line connecting the sBH and the SMBH and the orientation of the star's orbit relative to the sBH (see the \textit{left} panel of Figure~\ref{fig:schematic1}). 

The time $t_{\rm RL}$ it takes for the most energetic debris to reach the RL can be estimated numerically. $t_{\rm RL}$ is generally longer than the peak fallback time of the most bound debris (relative to the sBH) except for very small $R$ and $\Mbh$ (e.g., $R<10^{-4}\pc$ and $\Mbh\simeq 10^{5}\Msol$). This means it is plausible that the first TDE-like flare from the sBH can be generated while the most unbound ejecta is rushing towards the surface of RL of the sBH (\textit{middle} panel of Figure~\ref{fig:schematic1}).

Once the unbound debris reaches the RL boundary, the subsequent orbit would be governed by the potential of the SMBH. Figure~\ref{fig:orbit_hyperbolic} shows the eccentricity $e$ (\textit{left}) and  pericenter distance $r_{\rm p}$ (\textit{right}) of ``spilled'' debris relative to the $\Mbh=10^{6}\Msol$ SMBH, as a function of $R$ and $\theta$. The most energetic spilled debris produced in sBH-TDEs at $R\lesssim 10^{-3}-10^{-2}\pc$ is bound to the SMBH whereas the debris in sBH-TDEs at larger distances is unbound from the SMBH (white background). To zeroth order, the critical distance $\widehat{R}$ demarcating regions for the bound and unbound debris can be estimated by comparing the potential at the location of the sBH $E_{\mathrm{grav}}=-G\Mbh/R$ and the energy of the spilled debris $\Delta \epsilon$ (relative to the sBH). If $|E_{\mathrm{grav}}|\gg \Delta \epsilon$, the spilled debris does not have kinetic energy large enough to escape from the SMBH. However, if $|E_{\mathrm{grav}}|\ll \Delta \epsilon$, the spilled debris would escape. It follows that the critical distance is roughly,
\begin{align}\label{eq:maxR}
    \widehat{R}&\simeq \frac{G\Mbh}{\Delta\epsilon}\simeq 0.01\pc \left(\frac{\Mbh}{10^{6}\Msol}\right)\nonumber\\
    &\times\left(\frac{\mbh}{10\Msol}\right)^{-1/3}\left(\frac{\ms}{1\Msol}\right)^{-2/3}\left(\frac{\rs}{1\Rsol}\right),
\end{align}
which suggests that $\widehat{R}$ only depends on $\Mbh$ for given $\theta$, $\mbh$, and $\ms$. We confirm that the distributions of $e$ and $r_{\rm p}$ shown in Figure~\ref{fig:orbit_hyperbolic} shift horizontally while maintaining their dependence on $\theta$.  

The comparison between $|E_{\mathrm{grav}}|$ and $\Delta\epsilon$ allows us to understand the distribution of $e$ better (\textit{left} panel of Figure~\ref{fig:orbit_hyperbolic}). The debris produced closer to the SMBH tends to be more circular because the motion of the spilled debris is almost the same as that of the sBH given $\Delta\epsilon \ll |E_{\mathrm{grav}}|$. As shown in the \textit{right} panel, highly eccentric debris that can approach very close to the SMBH is mostly generated near $\widehat{R}$ where $\Delta \epsilon \simeq |E_{\mathrm{grav}}|$.

\begin{figure*}
    \centering
    \includegraphics[width=8.5cm]{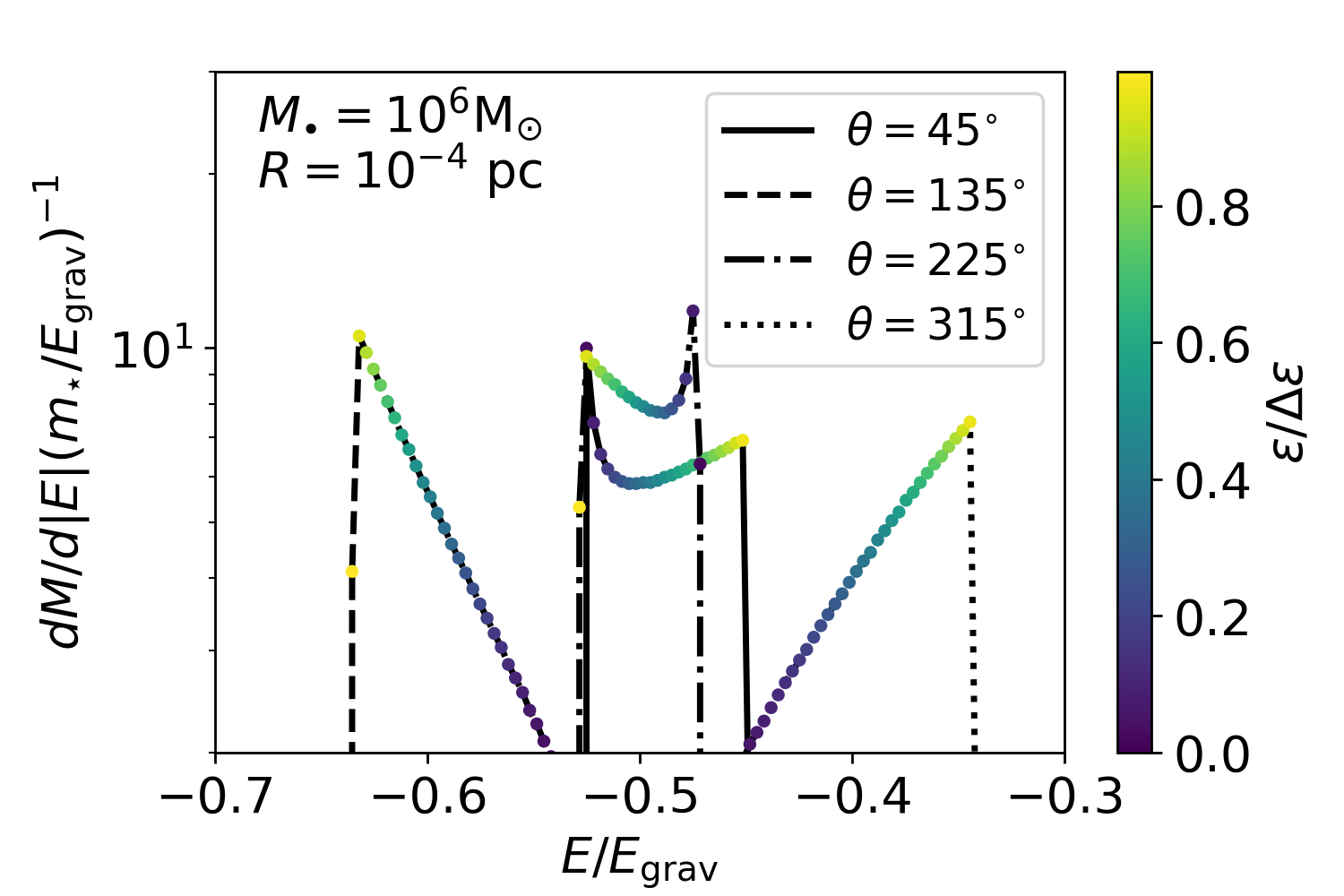}   
    \includegraphics[width=8.5cm]{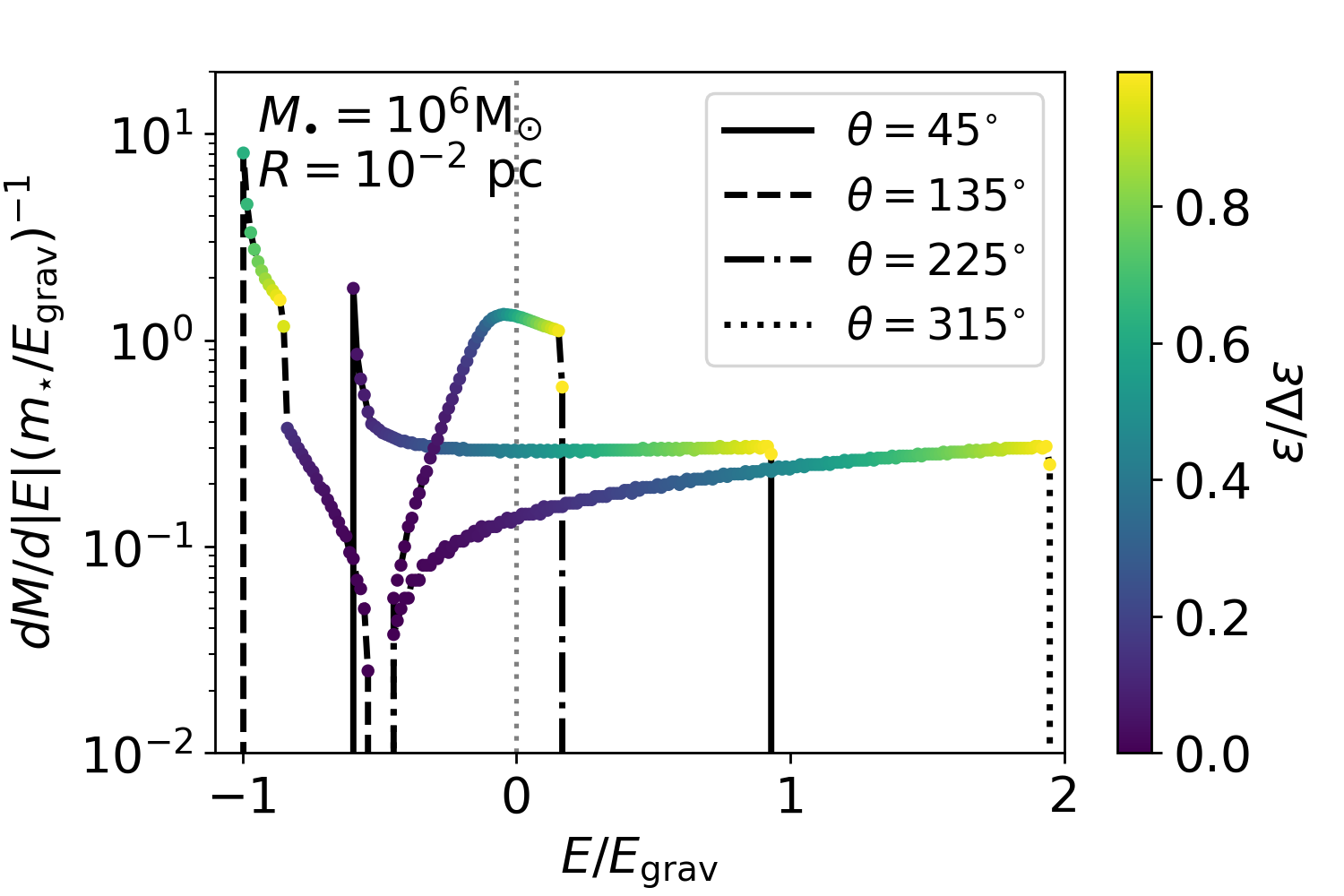}    
    \caption{\label{fig:dmde} The energy distribution $dM/dE$ of debris, originally unbound from the sBH, in the frame of the $10^{6}\Msol$ SMBH. $E_{\rm grav}$ is the local gravitational potential, $G\Mbh/R$.  The colors show how much the debris was unbound from the sBH originally: most unbound (yellow) to least unbound (blue). }
\end{figure*}

\begin{figure}
    \centering
    \includegraphics[width=9cm]{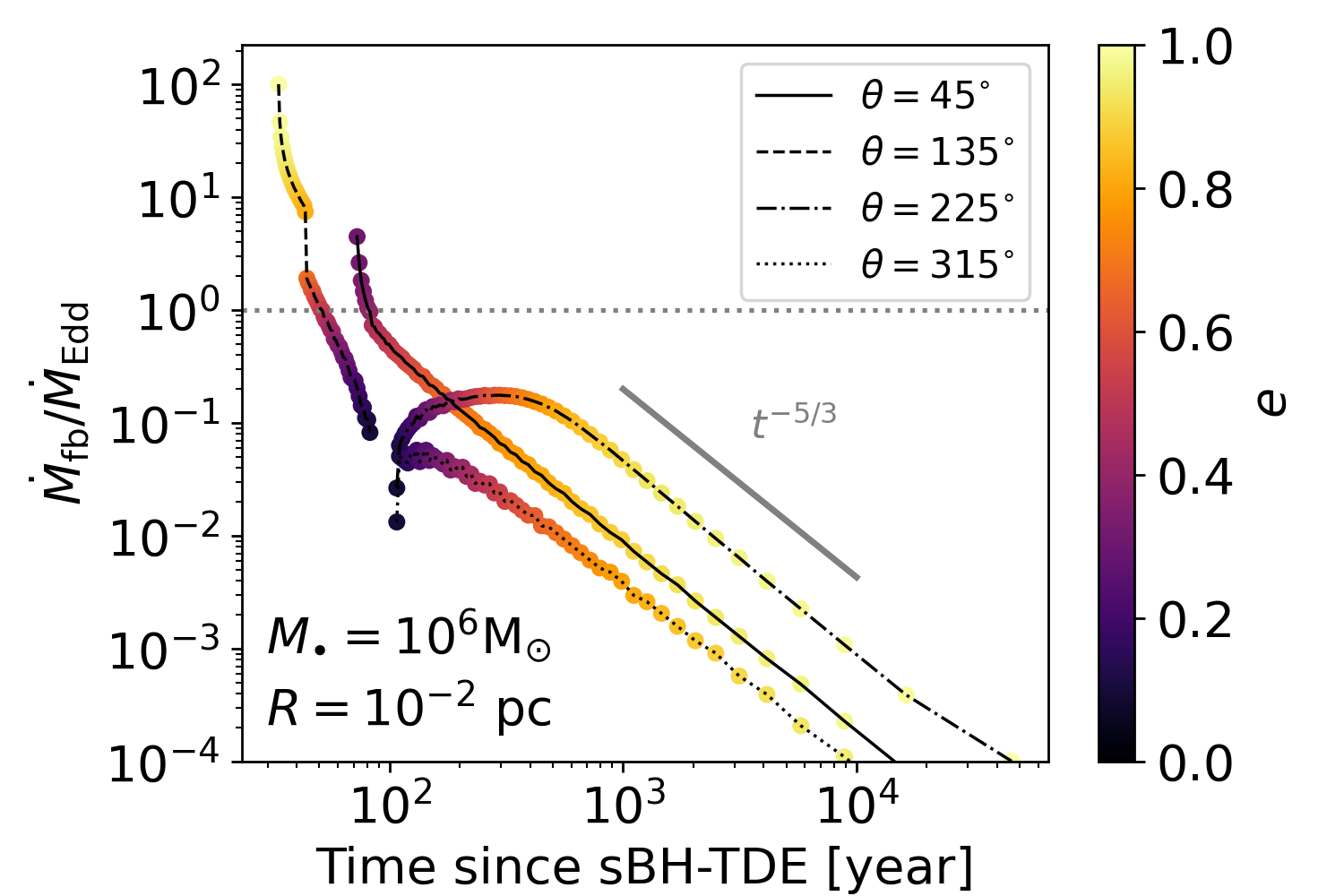}  
    \caption{\label{fig:fallback} The returning rate of spilled debris, originally unbound from the sBH following a sBH-TDE at distance $R$, to the SMBH for a range of values of $\theta$. $\dot{M}_{\rm Edd}= L_{\mathrm{Edd}}/c^{2}$ where $L_{\mathrm{Edd}}$ is the Eddington luminosity. The lines are color-coded based on the mass-weighted average eccentricity. Unlike ordinary TDEs, the fallback rates can 
 have a variety of shapes, depending on $\theta$, $R$, and $\Mbh$, including the $t^{-5/3}$ power-law decay.}
\end{figure}

\section{Energy distribution, Fallback rates and electromagnetic signatures}\label{sec:fallbackrate}

\subsection{Energy distribution}

Now we extend our analysis to the entire spilled debris. Here, we assume a top-hat energy distribution ($dM/d\epsilon =\ms/ 2\Delta\epsilon$) of originally unbound debris relative to the sBH \citep{Rees1988}. The energy distribution $dM/dE$ of the spilled debris in the frame of the $10^{6}\Msol$ SMBH is diverse and depends on $\theta$, $R$, and $\Mbh$. Here, $E$ represents the orbital energy relative to the SMBH. As an example, we show in Figure~\ref{fig:dmde} $dM/dE$ when a sBH-TDE happens at $R=10^{-4}\pc$ (\textit{left}) and $10^{-2}\pc$ (\textit{right}).  At $R=10^{-4}\pc$, $\Delta \epsilon\simeq E_{\rm grav}$ (more precisely $\Delta \epsilon=0.44E_{\rm grav}$). Consequently, while the energy distributions are concentrated near $E=0.5E_{\rm grav}$, corresponding to the circular orbit of the sBH, the angle dependence spreads the distributions around $0.5E_{\rm grav}$. The maximum range of $E/E_{\rm grav}$ would roughly be $-0.5E_{\rm grav}\pm \Delta \epsilon$. The eccentricity of the debris' orbit around the SMBH is 0.1 - 0.25. 

On the other hand, at $R=10^{-2}\pc$,  $\Delta \epsilon \gg E_{\rm grav}$. As a result, as shown in the \textit{right} panel of Figure~\ref{fig:dmde}, the energy distribution has a stronger $\theta-$dependence and it can spread over a wide range of $E$. The eccentricity of the debris's orbit relative to the SMBH in this case ranges from $0.1 - 1$, with a tendency of less bound debris (to the SMBH) on more eccentric orbits except for $\theta = 45^{\circ}$. This suggests that while more energetic debris, originally more unbound debris from the sBH, escapes the RL of the sBH at an earlier time, it would take a longer time for the debris to reach its pericenter on a more eccentric orbit around the SMBH.

As expected from the relation between $\Delta \epsilon$ and $E_{\rm grav}$, the energy distribution is degenerate between $\Mbh$ and $R$, for given $\mbh$ and $\ms$. In other words, the energy distribution $dM/d(E/E_{\rm grav})$ would look identical for the same $\Mbh/R$. For example, the distribution shown in the \textit{left} panel of the Figure~\ref{fig:dmde} is the same as that for $\Mbh=10^{8}\Msol$ and $R=10^{-2}\pc$.

\subsection{Fallback rates}

We compute the fallback rate under the common 'frozen-in' approximation for TDEs by SMBHs \citep{Rees1988}, where the debris is assumed to follow ballistic orbits following impulsive destruction of the star at the pericenter. Similarly, in our case, we assume that the spilled debris outside the RL of the sBH follows a ballistic orbit in the gravitational potential of the SMBH. The fallback rate may then be estimated as
\begin{equation}
\dot{M}_{\rm fb} = \frac{dM}{dt} = \frac{dM}{dE}\left|\frac{dE}{dt}\right|\,,
\label{eq:Mdot}
\end{equation}
where $dE/dt$ is computed as the analytic derivative of the energy with respect to the period, assuming ballistic orbits. The fallback rate is displayed in Fig.~\ref{fig:fallback} for sBH-TDEs at $R=10^{-2}\pc$ from the $10^{6}\Msol$ SMBH (see the \textit{right} panel of Figure~\ref{fig:dmde} for $dM/dE$). 
Depending on $\theta$ (also $R$ and $\Mbh$), the shape of the mass return curves varies. This is a clear difference from ordinary TDEs, where fallback curves are generally described by a rapid rise and power-law decay following a $t^{-5/3}$ power-law. In some of our cases, the fallback rate curves follows the $t^{-5/3}$ power-law (e.g., $\theta = 225^{\circ}$) where $dM/dE$ becomes flat. However, even in those cases, the slopes of the rising curves and the shape at the peak are different from each other. 

The peak mass return rate can be super-Eddington, and the mass return time is roughly of the order of the free-fall time at $R$. Given the scaling relation of $dM/dE$ with $\Mbh$ and $R$, the fallback time scales as $\propto \Mbh$ and $\dot{M}/\dot{M}_{\mathrm{Edd}}\propto \Mbh^{-2}$ for a given $\Mbh/R$. In other words, for $\Mbh=10^{8}\Msol$ and $R=1\pc$, the overall shape of the fallback curves is the same as that shown in Figure~\ref{fig:fallback}, while the rate decreases by four orders of magnitude and the return time increases by a factor of 100. So the mass return rates for that case become sub-Eddington, and the return times become longer than 3000 years.

Another difference from ordinary TDEs is that debris has a wide range of eccentricity and, therefore, pericenter. In these particular cases shown in Figure~\ref{fig:fallback}, except for $\theta = 45^{\circ}$, the spilled debris, originally less unbound from the sBH ($\epsilon\simeq 0$), becomes more tightly bound to the SMBH, reaching the pericenter on a relatively low-$e$ orbit at earlier times. As a result, the peak mass return rate corresponds to the return of low-$e$, tightly bound debris. 

\subsection{Electromagnetic signatures}

Broadly speaking, the events are characterized  by two flares, separated by a time delay. 
The first electromagnetic signature is produced by the tidal disruption of the star by the sBH. These sBH-TDEs \citep{Perets2016}
have been studied via numerical simulations with sBHs in isolation as well as in binaries \citep{Kremer2019,Lopez2019,Wang2021,Ryu2022,Kremer2022,Ryu2023a,Ryu2023b,Ryu2024}. 
Unlike the case of TDEs around SMBHs, cooling in the debris is found to be inefficient, resulting in the formation of an optically thick remnant disk. The low mass of the sBHs results in accretion rates
(taken as a proxy of the fallback rates) which are very high, possibly exceeding the Eddington luminosity by many orders of magnitude \citep[e.g.]{Wang2021,Kremer2019}. Under these conditions, strong wind/outflows are likely to develop \citep{Sadowski2014} and, in combination with the high accretion-induced spin, likely to launch a jet \citep{Krolik2012}. These conditions are resembling of those leading to the formation of gamma-ray bursts (GRBs), and in fact, sBHs have been proposed as possible progenitors for the subclass of ultra long GRBs \citep{Levan2014}, characterized by bright $\gamma / X$-ray flares of durations $\sim 10^3-10^4$~sec and luminosities generally $\sim 10^{48}-10^{49}$~erg/sec but occasionally as low as $\sim 10^{46}$~erg/sec\footnote{Also of possible interest is the transient Swift J1644+57 \citep{Levan2011}, which occurred in the nucleus of a galaxy, and which was interpreted both as a TDE \citep{Bloom2011} as well as an ultra-long GRB \citep{Quataert2012}.}. Reprocessing of the high-energy radiation within the outflow could also give rise to optical luminosity \citep{Kremer2019}.

This first flare is followed by the second flare once the sBH-unbound debris starts to fallback onto the SMBH.
This timescale can be evinced by 
Figure~\ref{fig:fallback}, here explicitly shown for the case of $\Mbh=10^6 \Msol$.
It is evident that the rate at which the debris rains back is highly dependent on the angle at which the plunge begins. 

Once fallback begins, the subsequent detailed evolution 
of the gaseous debris is largely determined by its ability to circularize in a disc. In contrast to the case of standard SMBH-TDEs, where the debris are typically highly eccentric and circularization requires a significant amount of energy dissipation through a combination of effects (such as thermal viscous and magnetic shears, compression of the stream at pericentre, general relativistic apsidal precession, see e.g. \citealt{Piran2015,Chen2018}), the spilled debris is dominated by low eccentricities (see the \textit{left} panel of Fig.~\ref{fig:orbit_hyperbolic}) for some cases, with the most bound material being the most circular one. Hence, the formation of a disk or a ring is likely to occur for the spilled debris.

\begin{figure*}
    \centering
    \includegraphics[width=8.5cm]{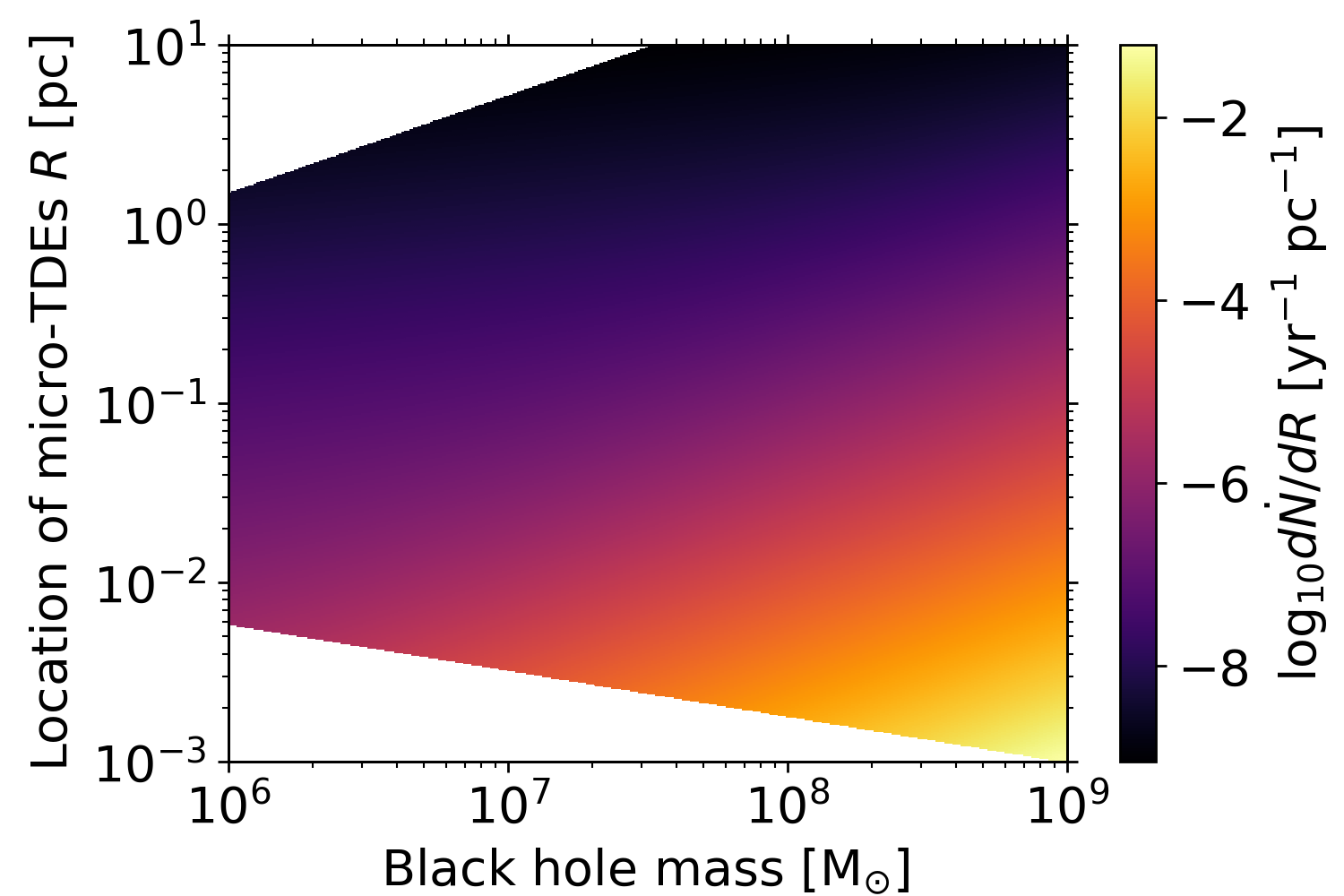}  
    \includegraphics[width=8.5cm]{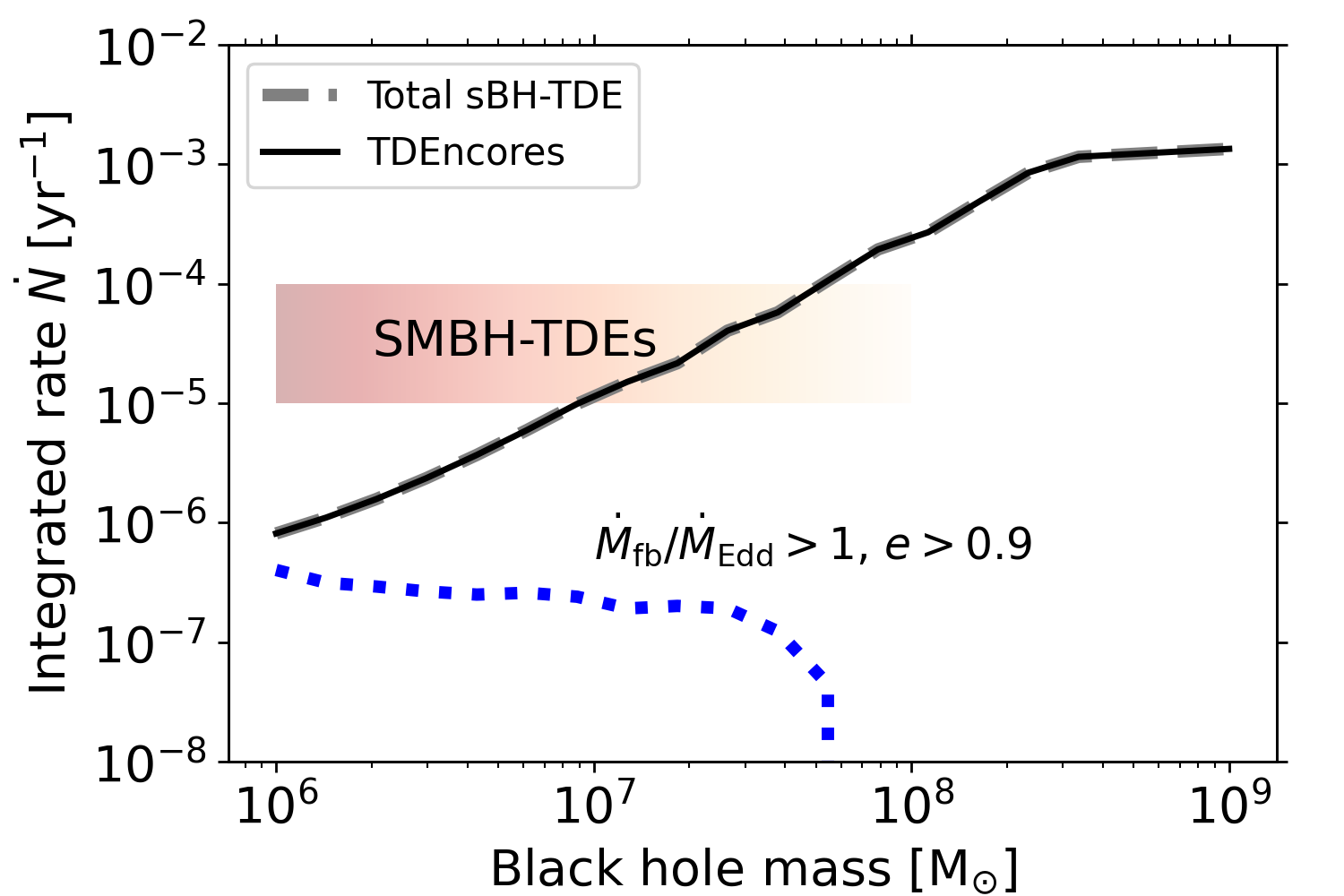}      
    \caption{\label{fig:rate} Differential rates of all sBH-TDEs of $1\Msol$ stars by $10\Msol$ BHs (\textit{left}) and integrated rates (\textit{right}). In the \textit{left} panel, the lower boundary is set by the maximum among the following distances: 1) $10r_{\rm g}$, 2) $r_{\rm t}$, and 3) $R$ above which there is at least one sBH and one star. In the \textit{right} panel, the grey solid line depicts the total rate for sBH-TDEs and the black dashed line the rate for sBH-TDEs 
 producing debris that could potentially fall into the SMBH and creating flares. Their rates are almost identical because most of sBH-TDEs happen close to the SMBH where the spilled debris is tightly bound to the SMBH. Note that the fallback rates of the spilled debris are super-Eddington in almost all sBH-TDEs. In addition, the rate of events with debris falling at a super-Eddington fallback rate on highly eccentric orbits ($e\gtrsim 0.9$) is depicted with the blue dotted line. The SMBH-'TDE'-like rates are indicated using the orange horizontal strip.}
\end{figure*}

Upon circularization, the subsequent evolution of a disk is set by the viscous timescale \citep{Shakura1973}. At high accretion rates, the flow is likely to be at least partly dominated by radiation-pressure \citep{Abramowicz1988}, which tends to make it puffier. The viscous timescale can be written as
\begin{align}
t_{\rm visc} &\simeq \frac{1}{\alpha} \left(\frac{R_{\rm circ}^3}{G \Mbh}\right)^{1/2} \left(\frac{H}{R_{\rm circ}}\right)^{-2}
\nonumber\\
&\simeq 1.3~{\rm yr}
\left(\frac{\alpha}{0.1}\right)^{-1}
\left(\frac{R_{\rm circ}}{10^3r_{\rm g,6}}\right)^{3/2}
\nonumber \\
&\times \left(\frac{\Mbh}{10^6\Msol}\right)^{-1/2}
\left(\frac{H/R_{\rm circ}}{0.2}\right)^{-2}
\,,
    \label{eq:tvisc}
\end{align}
where $\alpha$ is the viscosity parameter \citep{Shakura1973}, $R_{\rm circ}$ the circularization radius, $H$ the disk scale height,
and $r_{\rm g,6}$ is the gravitational radius of a $10^6M_\sun$ SMBH.
Note that, for an order of magnitude estimate, the above equation assumes the circularization radius to be at the pericenter of the spilled debris, but in reality there will be a spread of initial locations. As a reference, we normalized the equation using a moderate value for the disk scale height, however bearing in mind that there are likely going to be thickness variations across different events (e.g. \citealt{vanVelzen2019}).

Generally speaking, in the case where the circularization is efficient, if the viscous timescale is shorter than the fallback time, then the accretion rate onto the SMBH tracks the fallback rate. Conversely, if the viscous timescale is longer, then the accretion rate will be modulated by the viscous timescale.
Here we find the former situation to be true for most parameters, and hence we can assume to first approximation that the time evolution of the accretion rate $\dot{M}_{\rm acc}$ tracks that of the fallback rate. The time-delay between the high-energetic flare of the sBH-TDE and the peak of the second flare depends on a combination of parameters, as illustrated in Fig.~\ref{fig:fallback} and discussed in the previous subsection. The later-arrival flares are generally dimmer than the prompt arrivals, with a decay slope following the characteristic $t^{-5/3}$ law. 
The peak bolometric luminosity, $L_{\rm bol}=(\eta/0.1)\dot{M}_{\rm acc} c^2$ of these later flares is generally sub-Eddington, but can still reach luminosities compared with those of the standard TDEs.
In the specific example of Fig.~\ref{fig:fallback}, 
the peak bolometric luminosities for all the plunging angles, except for $\theta=135^o$, are in the range 
$L_{\rm bol}\sim (\eta/0.1) 5\times 10^{42}-5\times 10^{44}$~erg~s$^{-1}$. These luminosities, combined with the 'typical' TDE time-evolution and the very long delay $\gtrsim 100$~yr from the sBH-TDE flare, would make these events appear as 'impostors' in a galactic nucleus. 

The debris spilling from an angle $\theta=135^0$, on the other hand, being in a more radial trajectory towards the SMBH, have a much shorter fallback time, and can reach the SMBH on $\sim$~yr timescale following the micro flare. Note that the early debris are characterized by a high eccentricity, and hence some form of dissipation, as for the standard TDEs, is required for them to circularize.
A flare produced by these debris would be characterized by a rather high bolometric luminosity, approaching $\sim (\eta/0.1)  10^{45}$~erg~s$^{-1}$ for the example shown in Fig.~\ref{fig:fallback}. But what makes second flares of this kind especially stand out is their very steep time evolution, $\propto t^{-3}-t^{-4}$ in the early phase. 
While deviations from 
the classical $t^{-5/3}$ are generally expected in the optical also for standard TDEs \citep{Lodato2011}, such steep power laws are not generally predicted, and hence, when observed, could be indicative of a tidal disruption encore.
Measurements of the time delay between the sBH-TDE and the onset time of the second flare would  yield information on the location of the sBH within the NSC. 
Since shorter delay times correlate with closer sBHs, these types of events can provide a powerful diagnostics of the sBH population in the innermost region of NSCs.

\section{Rates}\label{sec:rate}

Let's consider a nuclear stellar cluster with a SMBH with mass $\Mbh$ at the center. We assume that the NSC extends out to the influence radius $r_{i}=G\Mbh/\sigma^{2}$ of the SMBH such that the enclosed mass is the same as $\Mbh$. Here, $\sigma$ is the velocity dispersion, for which we take the $\Mbh-\sigma$ relation by \citet{KormendyHo2013}. We also assume that the NSC consists of two stellar-mass objects, $\ms=1\Msol$ stars and $\mbh=10\Msol$ sBHs, both of which follow the Bahcall-Wolf profile \citep{BahcallWolf1976} such that the total number ratio of sBHs to stars is 0.01 \footnote{The mass ratio of the two populations would be also time-dependent. If stars are ejected at higher rates, the ratio could be even higher \citep{Gieles+2021}. }\citep[e.g.][]{Generozov+2018}. 

The differential rate $d\dot{N}/dR$ is estimated as,
\begin{align}
    \frac{d\dot{N}}{dR} = \frac{dN_{\bullet}}{dR} n_{\star}(R) \Sigma(R) v(R),
\end{align}
where $dN_{\bullet}/dR$ is the number of sBHs at $R$ within $R$ and $R+dR$, $n_{\star}$ the stellar number density, and $v$ the maximum of $\sqrt{G\Mbh/R}$ and $\sigma$. $\Sigma(R)$ is the encounter cross-section, estimated as,
\begin{align}
  \Sigma(R) = \pi r_{\rm t}^{2} \left( 1 + \frac{G m_{\bullet}}{r_{\rm t}v^{2}}\right),
\end{align}
where $r_{\rm t}=(\Mbh/\ms)^{1/3}\rs$ is the tidal disruption radius.

We first show the differential rate $d\dot{N}/dR$ as a function of $\Mbh$ and $R$ in the \textit{left} panel of Figure~\ref{fig:rate}. sBH-TDE rates are higher at smaller $R$ because of larger $dN_{\bullet}/dR$ and $n_{\star}$. In addition, at $R\lesssim 5\times10^{-3}\pc (\Mbh/10^{6}\Msol)$, $G\mbh/[r_{\rm t}v^{2}]<1$, $d\dot{N}/dR \propto v\propto R^{-1/2}$, indicating a larger $v$ near the SMBH also increases the rate. In addition, $dN_{\bullet}/dR$ increases with $\Mbh$: $dN_{\bullet}/dR\propto \Mbh^{0.3}$. Combining all these factors,  $d\dot{N}/dR\propto \Mbh^{1.14}$ for $R\ll5\times10^{-3}\pc (\Mbh/10^{6}\Msol)$ and $d\dot{N}/dR\propto \Mbh^{0.14}$ for $R\gg 5\times10^{-3}\pc (\Mbh/10^{6}\Msol)$

We estimate the total rate as $\dot{N}(\Mbh)=\int^{r_{i}}_{r_{0}}({\rm d}\dot{N}/{\rm d}R){\rm d}R$. Here, the integration lower $r_0$ limit is the maximum among the following distances: 1) $10r_{\rm g}$, 2) $r_{\rm t}$, and 3) $R$ above which there is at least one sBH and one star. For $\Mbh\lesssim$ a few $10^{8}\Msol$, $r_{0}$ is set by the distance at which $N_{\bullet}=1$ while it is set by $10r_{\rm g}$ for larger $\Mbh$ (see  Appendix~\ref{appendix:r0} and Figure~\ref{fig:r0} for the expressions of those distances and their values as a function of $\Mbh$). The total rate is depicted in the \textit{right} panel of Figure~\ref{fig:rate}. The total rate increases from $10^{-6}\yr^{-1}\pergal$ for $\Mbh\simeq 10^{6}\Msol$ to $\simeq 10^{-3}\yr^{-1}\pergal$ for $\Mbh\simeq 10^{9}\Msol$. The rate for events involving super-Eddington fallback (red line) is comparable to the total rate. Among those events, the rate of events involving super-Eddington fallback of debris on highly eccentric orbits (blue) is quite low: $\dot{N}\lesssim$ a few $10^{-7}\Msol$ for $\Mbh\lesssim$ $3\times 10^{7}\Msol$. 

Note that there are a few caveats in this estimate. First, NSCs and SMBHs coexist in almost all galaxies with mass $\lesssim 10^{9}-10^{10}\Msol$, corresponding to $\Mbh\lesssim 10^{8}\Msol$, whereas NSCs are less common towards the high galaxy mass end \citep{Neumayer2020}. Therefore, not every galaxy hosting a $\gtrsim 10^{9}\Msol$ SMBH would create sBH-TDEs as often as our rate estimate suggests. Second, stellar collisions are likely to be more frequent 
 than sBH-TDEs and could possibly be destructive \citep{Balberg+2013,Balberg+2023,AmaroSeoane2023,Rose+2023,Ryu+2024b}. Such destructive collisions can have a substantial impact on the stellar distribution in the vicinity of SMBHs, where sBH-TDEs are expected to be frequent. Consequently, destructive stellar collisions can significantly reduce the rate of sBH-TDEs.
 Lastly, we count all events where the closest approach distance of the star to the sBH is smaller than $r_{\rm t}$ as sBH-TDEs. However, in galactic nuclei, the relative orbits between the star and the sBH can be highly hyperbolic, potentially resulting in only a close fly-by rather than the disruption of the star. Therefore, our rates should be considered maximum rates.

\section{Summary and Conclusions} \label{sec:conclusion}

In this letter, we propose that TDEs by sBHs in NSCs can potentially generate two distinctive electromagnetic flares, one from the sBH and a subsequent one from the SMBH. For the
greatest majority of events, half the debris remains bound to the sBH, generating a first TDE-like flare, and the other half, originally unbound from the sBH, is still bound to the SMBH, and hence spills back to it, potentially generating a second flare.

Our key findings of our analytic model for the properties of the spilled debris are the following:
\begin{itemize}
    \item The originally bound debris to the sBH would evolve in the same way as in sBH-TDEs by sBHs in isolation, generating the first flare on the fallback timescale of $\simeq O(1)$ days. The originally unbound debris from the sBH would be spilled from its Roche Lobe. For sBH-TDEs at $R\lesssim 0.01\pc (\Mbh/10^{6}\Msol)$ (Equation~\ref{eq:maxR}), even the most energetic ``spilled" debris remains bound to the SMBH, falling into it and potentially creating the second flare on a time scale of $\gtrsim O(1)$ yrs. 
    
    \item Unlike for standard TDEs, the spilled debris that remains bound to the SMBH, can have a wide range of energy and angular momentum, depending on various factors including the initial angle of departure of the debris relative to the motion of the sBH ($\theta$), SMBH mass ($\Mbh$), the location of sBH-TDEs ($R$). As a consequence, the shape of the fallback rate curves varies (see Figure~\ref{fig:fallback}). For instance, the peak return rates can be both super-Eddington and sub-Eddington for given $\Mbh$ and $R$, depending on $\theta$. Also, it is possible that debris on a less circular orbit can have shorter delays and higher return rates.  
    
    \item The electromagnetic signatures are characterized by a double flare: a short, $\gamma$-ray/X-ray bright one, possibly resembling a very long GRB, followed by a dimmer second one, with 
    bolometric luminosity consistent with 
    the bulk of standard TDEs. Time delays vary from a few years to hundreds of years. Second flares with very long delays generally display the characteristic $\propto t^{-5/3}$, which      
    would make these appear as 'impostors' in the standard TDE population. Early and bright arrivals, on the other hand, are characterized by much steeper decays.
    
    \item The total rates of close encounters potentially resulting in sBH-TDEs with a secondary flare range from $10^{-6}-10^{-3} \yr^{-1}\pergal$ for $\Mbh\simeq 10^{6}-10^{9}\Msol$.
    In almost all these events, the fallback rates can be super-Eddington.  However, the rate of events creating spilled debris falling back to the SMBH at super-Eddington rates on highly eccentric orbits is $\lesssim 10^{-7}\yr^{-1}\pergal$, independent of the $\Mbh$. 
    
\end{itemize}

Our newly discovered astrophysical phenomenon carries a broad range of implications. When a double peak is observed, either in real-time or via an archive search of positionally-coincident X-ray/$\gamma$-ray transients, information on the position of the sBH  can be gained. 
More generally, rates of the double flare events can help calibrate the number of sBHs in the NSC.
Among other, this is important for the AGN channel of BH-BH mergers \citep{McKernan2020} since a large fraction of BHs is expected to be captured into the AGN disk from the NSC \citep{Fabj2020,Wang2024}.

Double flares with short time-delays, associated with sBHs in tight orbits, could be seen during the Extreme Mass Ratio Inspiral (EMRI) of the LISA band if a star happens to be within the innermost several $r_{\rm g}$ of the SMBH \citep{Derdzinski2021} when a sBH spirals in.

For events in which the first sBH-TDE flare is missed, the second flare from the SMBH can appear as an 'impostor' among the standard TDEs from the SMBH. Interestingly, a fraction of the observed TDEs displays slopes which deviate from the canonical expectations for TDEs \citep{Gezari2021,Hammerstein+2023}

The implications of these impostors can be particularly significant if a 'TDE'-like event
is observed around a very massive SMBH.
TDE rates are expected to drop to zero above $\sim 10^8 \Msol$ for a non-rotating SMBH , but only fade at around  
$7\times 10^8 M_\sun$ for maximally spinning SMBHs (e.g. \citealt{Kesden2012}). 
Hence TDE rates have diagnostic power to measure SMBH spins, and even constrain a subclass of axions \citep{Du2022}. Impostors
in this range of mass can affect spin measurements.

Since this has been the very first paper (to the best of our knowledge)  studying the occurence and properties of sBH-TDE/SMBH-TDE coupled events,
we have kept our basic set up and assumptions simple, as we remind in the following. 
 First, we assume that the orbit of the original star is parabolic. However, in galactic nuclei, the orbit can be highly hyperbolic. If hyperbolic sBH-TDEs occur, the energy distribution of debris would shift towards higher energy. This change can affect $\widehat{R}$ and the subsequent evolution of debris around the SMBH. Second, we consider the sBH to be in a circular orbit around the SMBH. Deviations from the circular orbit would also affect the evolution of the spilled debris because its energy and angular momentum at the RL of the sBH would differ, depending on the orbital phase of the sBH. For instance, in the case with the sBH on an eccentric orbit, the orbit of spilled debris can be eccentric (relative to the SMBH), even in sBH-TDEs that happen very close to the SMBH, as opposed to the case with a circular sBH (see the \textit{left} panel of Figure~\ref{fig:orbit_hyperbolic}). Third, we consider coplanar cases where the star, sBH, and SMBH orbit in the same plane for simplicity, which may be a good approximation for AGN. In gas-free environments, the star's orbit relative to the sBH is not necessarily aligned with the sBH-SMBH orbit. The additional angle required to describe a non-zero mutual inclination would introduce a more complicated angle dependence. Fourth, we assume that orbit of debris is ballistic around one of the two BHs. To zeroth order, this approximation holds when considering the orbital evolution of debris in a system involving a large mass contrast between the sBH and the SMBH. Lastly, we assume sBH-TDEs in gas-free galaxies. However SMBHs tend to be surrounded by a gas medium \citep[e.g., our Galactic center, ][]{Gillessen+2019}. In such cases, interactions of the infalling debris with surrounding gas --particularly material falling from a long distance -- can affect the orbit of the debris and also lead to debris dissolution via Kelvin-Helmholtz instability \citep[e.g., see ][]{Bonnerot+2016,Ryu+2024c}. To accurately understand the observables and orbital evolution of debris in various environments, detailed hydrodynamics simulations with proper gravity calculations from both BHs are required, which we leave for our follow-up projects.

\section*{Acknowledgments}

The authors are grateful to the anonymous referee for constructive comments and suggestions. We thank Philip Armitage and Zoltan Haiman for helpful discussions. R.P. gratefully acknowledges support by NSF award AST-2006839. The Center
for Computational Astrophysics at the Flatiron Institute is supported by the Simons Foundation.

\software{
matplotlib \citep{Hunter:2007}
}
%\bibliography{biblio}

\appendix
\section{Scaling relations}\label{appendix:r0}

We consider a cluster with mass $M_{\rm NSC}$ consisting of objects with mass $m$ and radius $r$ following a power-law profile,
\begin{align}
    \rho(R) = \rho_{i} \left(\frac{R}{r_{i}}\right)^{-\gamma}, 
\end{align}
where $r_{i}$ is the influence radius and $\rho_{i}$ is the density at $R=r_{i}$.
Although we consider a Bahcall-Wolf profile with $\gamma=7/4$ in this paper, we leave $\gamma$ as a free parameter in this analysis. Taking the $\Mbh-\sigma$ relation from \citet{KormendyHo2013},
\begin{align}
    \sigma = 200 ~{\rm km~s}^{-1} \left(\frac{\Mbh}{0.309\times 10^{9}\Msol}\right)^{1/4.38},
\end{align}
the influence radius $r_{i}$ is,
\begin{align}
    r_{i} = \frac{G\Mbh}{\sigma^{2}} \simeq 1.5\pc \left(\frac{\Mbh}{10^{6}\Msol}\right)^{0.54}.
\end{align}
It follows that $\rho_{i}$, the enclosed mass of stars $M(<R)$, and differential mass $dM/dR$ are,
\begin{align}
    \rho_{i} &= \frac{(3-\gamma)M_{\rm NSC}}{4\pi r_{i}^{3}}\simeq 3.1\times10^{4}\Msol\pc^{-3}\left(\frac{M_{\rm NSC}}{10^{6}\Msol}\right)\left(\frac{\Mbh}{10^{6}\Msol}\right)^{-1.63},\\
    M(<R)   & = \frac{4\pi \rho_{i} r_{i}^\gamma}{(3-\gamma)}R^{3-\gamma} \simeq 2000\Msol\left(\frac{M_{\rm NSC}}{10^{6}\Msol}\right) \left(\frac{\Mbh}{10^{6}\Msol}\right)^{-1.63+0.54\gamma}\left(\frac{R}{10^{-2}\pc}\right)^{3-\gamma},\\
    \frac{dM}{dR} & = 4 \pi R^{2} \rho(R) = 2.4\times 10^{5}\Msol \pc^{-1} \left(\frac{M_{\rm NSC}}{10^{6}\Msol}\right)\left(\frac{\Mbh}{10^{6}\Msol}\right)^{(-1.63+0.54\gamma)}\left(\frac{R}{10^{-2}\pc}\right)^{2-\gamma}.
\end{align}

The minimum distance at which there is at least one cluster member object, or $M(<R)/m=N=1$, is,
\begin{align}
    R(N=1) = \left(\frac{(3-\gamma)m}{4\pi \rho_{i} r_{i}^\gamma}\right)^{1/(3-\gamma)}\simeq 2\times10^{-5}\pc \left(\frac{M_{\rm NSC}}{10^{6}\Msol}\right)^{-1/(3-\gamma)}\left(\frac{\Mbh}{10^{6}\Msol}\right)^{(1.63-0.54\gamma)/(3-\gamma)}\left(\frac{m}{1\Msol}\right)^{1/(3-\gamma)}.
\end{align}

\begin{figure}
    \centering
    \includegraphics[width=18cm]{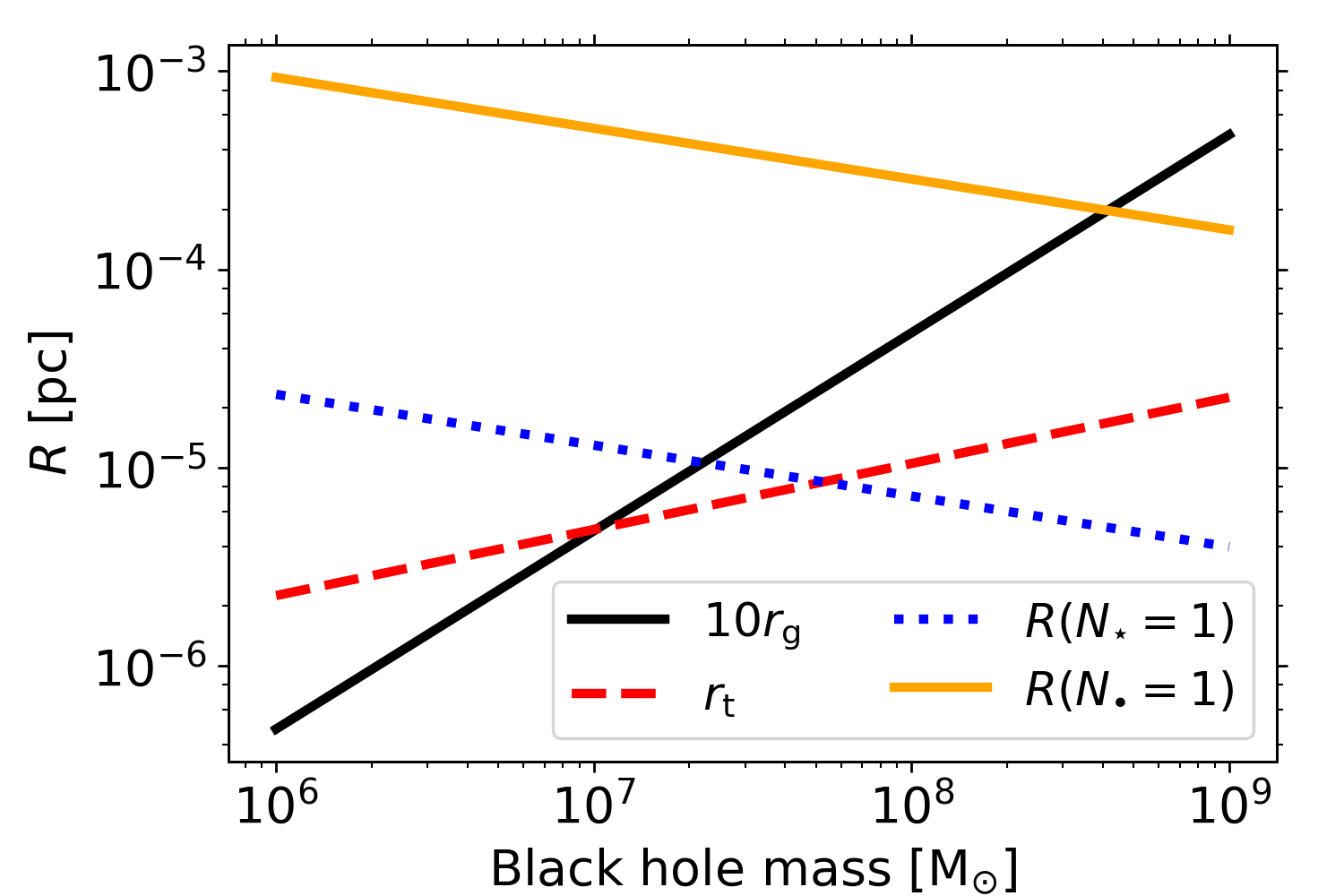}     
    \caption{\label{fig:r0} Several distances that determines the lower limit in the integration for the event rate: $10r_{\rm g}$ (black solid), tidal radius $r_{\rm t}$ (red dashed), the minimum distance at which there is at least one star $R(N_{\star}=1)$ (blue dashed), and the minimum distance at which there is at least one sBH $R(N_{\bullet}=1)$ (orange solid). The maximum of those distances is used as the integration lower limit for the total rate calculation. }
\end{figure}

In Figure~\ref{fig:r0}, we estimate several characteristic distances, whose maximum is used as the integration lower limit $r_{i}$ to estimate the total rate $\dot{N}$. The characteristic distances are $10r_{\rm g}$ (black solid), $r_{\rm t}$ (red dashed), $R(N_{\star}=1)$ (blue dashed), $R(N_{\bullet}=1)$ (orange solid), $\bar{R}(r_{\rm t})$ (cyan dotted). $r_{i} = R(N_{\bullet}=1)$ for $\Mbh\leq 4\times 10^{8}\Msol$ and $r_{i}=10r_{\rm g}$ $\Mbh\geq 4\times 10^{8}\Msol$. The distance at the transition determines the location of the knee in the \textit{right} panel of Figure~\ref{fig:rate}.

\end{document}